\documentclass[final,3p,times]{elsarticle}
\usepackage{}
\usepackage[dvipsnames]{xcolor}

\usepackage{amsmath}
\usepackage{amssymb}

\makeatletter
\newcommand{\rmnum}[1]{\romannumeral #1}
\newcommand{\Rmnum}[1]{\expandafter\@slowromancap\romannumeral #1@}
\makeatother

\journal{Physica A}

\begin{document}

\begin{frontmatter}

{\footnotesize{
\noindent Revised version\\
Original article: {\color{Cyan}Physica A 444 (2016) 928--939}, 
DOI: {\color{Cyan}http://dx.doi.org/10.1016/j.physa.2015.10.048}\\
Corrigendum: {\color{Cyan}Physica A 447 (2016) 569--570}, 
DOI: {\color{Cyan}http://dx.doi.org/10.1016/j.physa.2015.12.010}\\
\\
\\}
}



\title{A pathway-based network analysis of hypertension-related genes}


\author[BJUAS1,YNU1]{Huan Wang}\ead{hwang227@126.com}

\author[YNU1,BJUAS2]{Jing-Bo Hu}

\author[YNU1]{Chuan-Yun Xu}

\author[YNU2]{De-Hai Zhang}

\author[CXNU]{Qian Yan}

\author[YNU1,KLU]{\\Ming Xu}

\author[YNU1]{Ke-Fei Cao\corref{cor1}}\ead{kfcao163@163.com}

\author[PHE,IC]{Xu-Sheng Zhang}\ead{xu-sheng.zhang@phe.gov.uk}

\address[BJUAS1]{School of Computer Science and Technology, 
Baoji University of Arts and Sciences, 
Baoji, Shaanxi 721016, China}

\address[YNU1]{Center for Nonlinear Complex Systems, 
Department of Physics, 
School of Physics Science and Technology, 
Yunnan University,\\
Kunming, Yunnan 650091, China}

\address[BJUAS2]{School of Electronic and Electrical Engineering, 
Baoji University of Arts and Sciences, 
Baoji, Shaanxi 721016, China}

\address[YNU2]{School of Software, 
Yunnan University, 
Kunming, Yunnan 650091, China}

\address[CXNU]{School of Physics and Electronic Science, 
Chuxiong Normal University, 
Chuxiong, Yunnan 675000, China}

\address[KLU]{School of Mathematical Sciences, 
Kaili University, 
Kaili, Guizhou 556011, China}

\address[PHE]{Modelling and Economics Unit, 
Centre for Infectious Disease Surveillance and Control, 
Public Health England,\\
61 Colindale Avenue, London NW9 5EQ, UK}

\address[IC]{Medical Research Council Centre for Outbreak Analysis and Modelling, 
Department of Infectious Disease Epidemiology,\\
School of Public Health, 
Imperial College London, 
Norfolk Place, London W2 1PG, UK}

\cortext[cor1]{Corresponding author. 
Tel.: +86 871 65031605.}

\begin{abstract}
Complex network approach has become an effective way to describe 
interrelationships among large amounts of biological data, which is 
especially useful in finding core functions and global behavior of 
biological systems. Hypertension is a complex disease caused by many 
reasons including genetic, physiological, psychological and even 
social factors. In this paper, based on the information of biological 
pathways, we construct a network model of hypertension-related genes 
of the salt-sensitive rat to explore the interrelationship between 
genes. Statistical and topological characteristics show that the 
network has the small-world but not scale-free property, and 
exhibits a modular structure, revealing compact and complex 
connections among these genes. By the threshold of integrated 
centrality larger than $0.71$, seven key hub genes are found: 
{\it Jun}, {\it Rps6kb1}, {\it Cycs}, {\it Creb3l2}, {\it Cdk4}, 
{\it Actg1} and {\it RT1-Da}. These genes should play an important 
role in hypertension, suggesting that the treatment of hypertension 
should focus on the combination of drugs on multiple genes.
\end{abstract}


\end{frontmatter}



\section{Introduction}
\label{sec:1}

The study of complex systems clearly shows that the global behavior 
of systems is determined by their structure rather than by the 
properties of their individual parts. The complex network approach 
has become a powerful tool for studying complex systems, and the 
global properties of systems are usually studied by abstracting 
individual elements of systems into nodes and reducing interactions 
between elements to edges between 
nodes~\cite{WS1998n,BA1999s,S2001n,N2003sr,WYY2006pre}. Such an 
approach has been widely applied to understanding gene functions in 
biological and medical 
research~\cite{JTA2000n,BO2004nrg,KH2007pa,TSP2009pa,D2011pa}.

Essential hypertension, which accounts for about $90\%$--$95\%$ of 
all cases of hypertension~\cite{CO2000c}, is a disease caused by 
long-term interaction between genetic and environmental factors, and 
salt is one of the important environmental factors~\cite{CNB2012h}. 
The blood pressure response to salt loading or salt restriction is 
heterogeneous among individuals, which is known as salt 
sensitivity~\cite{WFF2001h,A2002ije,RIV2007ndt}. Salt sensitivity is 
the genetic susceptibility of individual blood pressure response to 
salt, and is an intermediate phenotype of essential 
hypertension~\cite{R2000pr,C2006nrg}. The people who suffer the 
salt-sensitive (SS) hypertension account for about $50\%$ of 
hypertensive patients~\cite{RIV2007ndt}. Although the clinical 
research and treatment of hypertension have improved 
dramatically~\cite{CCDN2000c,WMCK2004h,SJJ2011chr}, its molecular 
mechanisms and pathologies involved are still difficult to ascertain.

Many omic data have been obtained and become available through 
advanced high-throughput technologies, which provide the basis for 
studying the relationship of biological data by network 
approach~\cite{BO2004nrg,D2011pa}. Various biomolecular networks 
have been constructed to discover essential functions and mechanisms 
of biological phenomena~\cite{JTA2000n,KH2007pa,TSP2009pa}. For 
instance, Censi et al. studied the gene regulatory networks induced 
in heart tissue by atrial fibrillation~\cite{CGBC2011itbe}. 
Demicheli and Coradini analyzed breast cancer behavior using gene 
regulatory networks~\cite{DC2011ao}. Therefore, it is of 
significance to understand hypertension disease at system level using 
the complex network approach.

In our previous study~\cite{WXHC2014pa}, we constructed a 
hypertension-related gene co-expression network by focusing on the 
analysis of gene expression data (GED)~\cite{LLW2008pg} among the 
Dahl SS rat~\cite{DHT1962jem,R1982h} and two consomic rat 
strains~\cite{D1998jh,CRJ2004jp}, where the $335$ nodes are 
individual genes and the connections are derived from the expression 
correlations. This is a theoretical analysis based on GED to 
determine the key hub genes (nodes) and explore the relationship 
between these hub genes and hypertension. However, to get more 
biologically relevant information about hypertension, a pathway-based 
gene network should also be constructed using the actual biological 
correlations.

In the present work, we attempt to study the genes that are involved 
in SS hypertension based on the information of biological pathways. 
A biological pathway is a series of actions among molecules in a cell 
that leads to a certain product or a change in a 
cell~\cite{K2002s,DGvH2003bb}. Such a pathway can trigger the 
assembly of new molecules or turn genes on and off. Since biological 
pathways such as metabolic and signal transduction pathways can 
directly be viewed as interconnected processes of molecular species 
in the cell~\cite{K2002s,DGvH2003bb}; therefore, constructing a 
pathway-based gene network could help to disentangle the actual 
biological interactions between genes. 

In this paper, we will construct the network model of 
hypertension-related genes according to whether these genes are 
involved in the same pathways in the KEGG\footnote{
KEGG (Kyoto Encyclopedia of Genes and Genomes) website: 
http://www.kegg.jp/ or http://www.genome.jp/kegg/.} 
database. Network approach will be employed to investigate the 
possible relations between network structure and hypertension-related 
genes based on these data. Through calculating several statistical 
indices and analyzing topological characteristics of the network, we 
find that the pathway-based gene network exhibits the small-world but 
not scale-free property. Meanwhile, the network also exhibits a 
modular structure: the nodes of the network can be properly divided 
into groups within which the nodes are highly connected, but between 
which they are much less connected. The modular structure analysis 
can visualize the weak connections of the network, and thus help us 
to study drug targets of hypertension. The results from this paper 
and the analysis in Ref.~\cite{WXHC2014pa} would complement each other.

The rest of this paper is organized as follows. In 
Section~\ref{sec:2}, we introduce the data source and construct the 
pathway-based gene network model of hypertension. In 
Section~\ref{sec:3}, we analyze the statistical and topological 
characteristics of the gene network. The modular structure of the 
network is presented in Section~\ref{sec:4}, while 
Section~\ref{sec:5} presents summary and concluding remarks.

\section{Data source and network construction}
\label{sec:2}

The Dahl SS rat, proposed by Dahl et al. in the early 
1960s~\cite{DHT1962jem,R1982h}, is a widely used genetic model of 
human hypertension. The consomic rat strains, used as the 
normotensive control for the Dahl SS rat, are generated by 
substituting a chromosome or a part of a chromosome from a normal 
rat strain for the corresponding genomic region of the SS 
rat~\cite{LLW2008pg,D1998jh,CRJ2004jp}. Previous research has 
shown that substitution of chromosome 13 or 18 can attenuate 
hypertension~\cite{JLG2005pg,LLW2008pg}. Our study will focus on the 
hypertension-related genes listed in Ref.~\cite{LLW2008pg} by 
analysis of biological pathways.

Let us consider an undirected network $G_{H}=(V_{H},E_{H})$, where 
$V_{H}=\{v_{i}\}$ $(i=1,2,\ldots,N)$ denotes the set of $N$ nodes, 
and $E_{H}=\{v_i,v_j\}$ the set of edges or connections between 
nodes. We will use the following notation: $A_{ij}=1$ indicates that 
there is an edge between nodes $v_{i}$ and $v_{j}$; and $A_{ij}=0$ 
otherwise. Our pathway-based gene network model is constructed in 
two steps.

{\it Step 1. Selection of genes according to pathways}

We first extract nodes from the $335$ different hypertension-related 
genes\footnote{
All these genes are given in Table S2 (Excel file) of Supplemental 
Figures and Tables of Ref.~\cite{LLW2008pg}, which can be readily 
accessed from the website: 
http://physiolgenomics.physiology.org/content/34/1/54/suppl/DC1.} 
given in Ref.~\cite{LLW2008pg} based on the KEGG PATHWAY Database. 
From this Database, we can easily obtain the information of whether 
a gene is involved in single or multiple pathways. Here, we only 
consider the pathways including hypertension-related genes, and the 
genes not involved in any pathway are excluded from our study. Thus, 
the $N=90$ hypertension-related genes, which are involved in $157$ 
pathways in the KEGG Database~\cite{W2013DScD}, are extracted to 
serve as nodes of the network model, where each node represents an 
individual gene shown as gene symbol or CloneID. Examples of ten 
genes involved in one or more pathways are shown in 
Table~\ref{gene.vs.pathway}; here, a pathway is represented by the 
entry name of the Database, called the KEGG object identifier 
consisting of a database-dependent prefix and a five-digit number 
(such as rno04066, where the prefix ``rno'' designates the species 
to be rat).

\begin{table}[t]
\centering
\scriptsize
\caption{Examples of the correspondence between genes and pathways.}
\label{gene.vs.pathway}
\begin{tabular}[c]{llrll}
\hline
Gene  & \qquad\qquad\qquad & Gene ID  & \qquad\qquad\qquad & Pathway(s)\\
\hline
{\it Timp1}     && 116{\,}510{\ } && rno04066\\
{\it Casp6}     &&  83{\,}584{\ } && rno04210\\
{\it Ank3}      && 361{\,}833{\ } && rno05205\\
{\it Aqp1}      &&  25{\,}240{\ } && rno04964, rno04976\\
{\it Kcnj1}     &&  24{\,}521{\ } && rno04960, rno04971\\
{\it Ctsd}      && 171{\,}293{\ } && rno04142, rno05152\\
{\it Hist1h2ai} && 502{\,}129{\ } && rno05034, rno05322\\
{\it Sdc1}      &&  25{\,}216{\ } && rno04512, rno04514, rno05144, rno05205\\
{\it Col4a1}    && 290{\,}905{\ } && rno04151, rno04510, rno04512, rno04974, rno05146, rno05200, rno05222\\
{\it Fzd2}      &&  64{\,}512{\ } && rno04310, rno04390, rno04916, rno05166, rno05200, rno05205, rno05217\\
\hline
\end{tabular}
\end{table}

{\it Step 2. Establishment of connections}

\begin{figure}[!]
\begin{center}
\includegraphics[width=7.5cm]{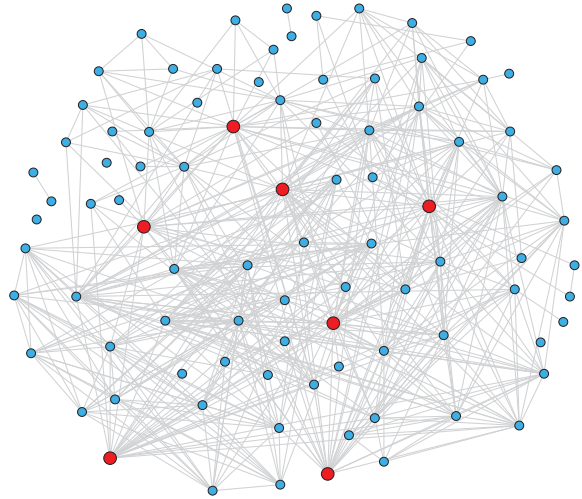}
\caption{Illustration of the pathway-based gene network of 
hypertension for the SS rat with all $90$ nodes and $482$ edges. 
The large red nodes represent the seven hub genes. 
(For interpretation of the references to color in this figure legend, 
the reader is referred to the web version of this article.)}
\label{gene network}
\end{center}
\end{figure}

We now consider the correlations between any two genes according to 
the information of pathways. If two genes $i$ and $j$ are involved in 
the same pathway(s), then a connection is made between such two genes 
(nodes): 
\begin{equation}
A_{ij}=\left \{ \begin{array}{lcl}
       1 & & \mbox{if genes}\  i \ \mbox{and}\  j \ \mbox{are in the same pathway(s)};\\
       0 & & \mbox{otherwise}.
       \label{adjacency matrix}
       \end{array} \right.
\end{equation}
Here $i,j=1,2,\ldots,90$ and $i\neq j$ (note that $A_{ii}=0$ because 
we do not consider self-connections of nodes, i.e., self-interactions 
of genes). Consequently, there are biological regulatory 
relationships between genes $i$ and $j$ when $A_{ij}=1$. In such a 
way, we have constructed the pathway-based network of 
hypertension-related genes, which contains $90$ nodes (genes) and 
$482$ edges (connections), as shown in Fig.~\ref{gene network}.

\section{Statistical and topological characteristics of the gene network}
\label{sec:3}

In this section, we analyze the pathway-based gene network of 
hypertension by calculating the following indices: degree (and 
average degree), degree distribution, average path length, 
clustering coefficient, assortativity coefficient, and four 
centrality indices (degree centrality, betweenness centrality, 
closeness centrality and integrated centrality), which can provide 
us with statistical and topological characteristics of the gene 
network.

\subsection{Degree and degree distribution}
\label{subsec:3.1}

The degree $k_{i}$ of a node $i$ is the number of edges connecting 
to the node. The average of $k_{i}$ over all nodes is called the 
average degree of the network, and is denoted as 
$\langle k\rangle =\frac{1}{N}\sum_{i=1}^{{N}} k_{i}$. 
The degree distribution function $P(k)$ gives the probability that a 
randomly selected node has exactly $k$ edges~\cite{S2001n,N2003sr}. 
The degree distribution is one of the most basic quantitative 
properties of a network.

\begin{figure}[!t]
\begin{minipage}[t]{0.5\linewidth}
\centering\hspace{0.0cm}
\includegraphics[width=8.05cm]{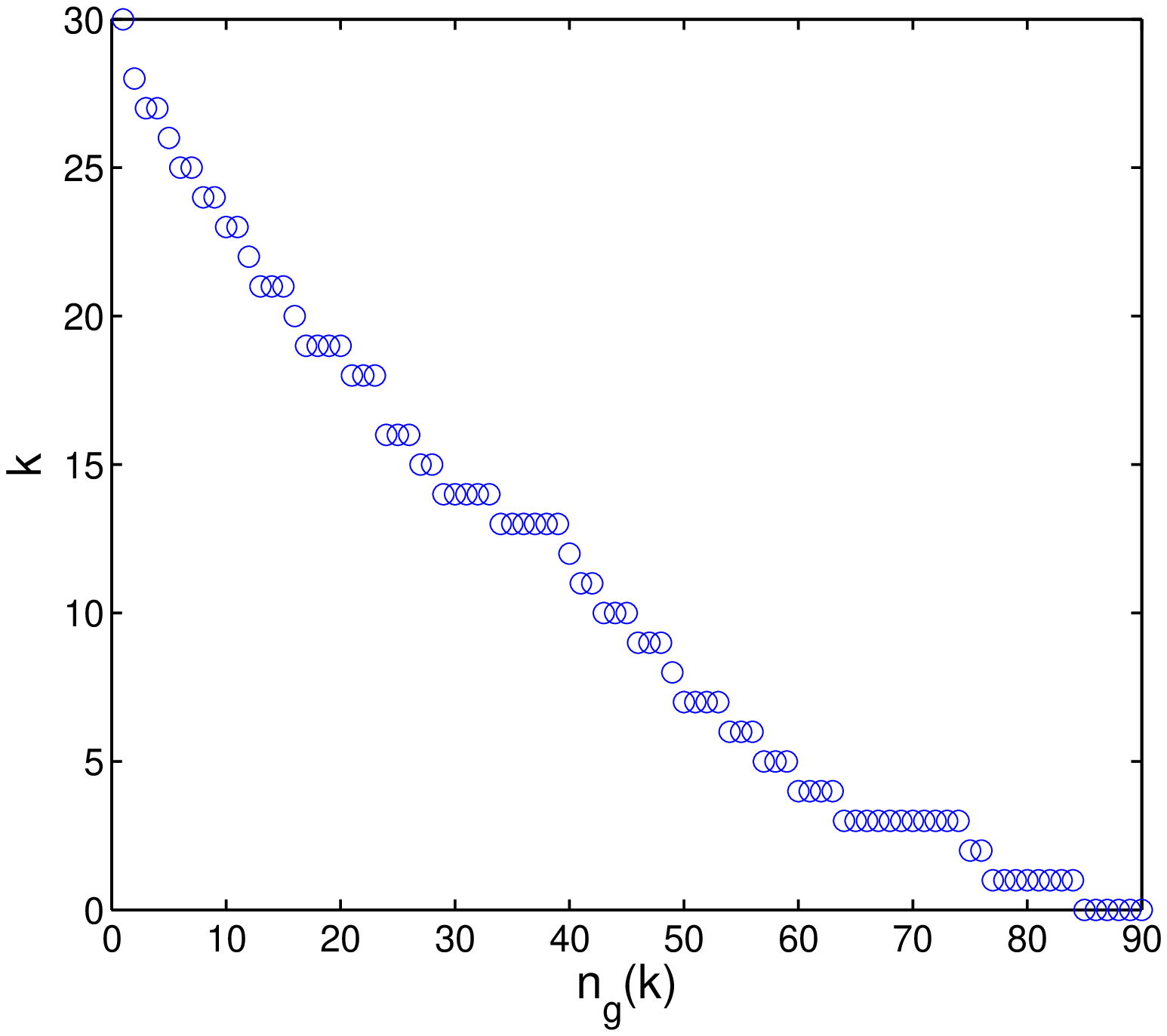}
\end{minipage}
\begin{minipage}[t]{0.5\linewidth}
\centering\hspace{-0.45cm}
\includegraphics[width=8.03cm]{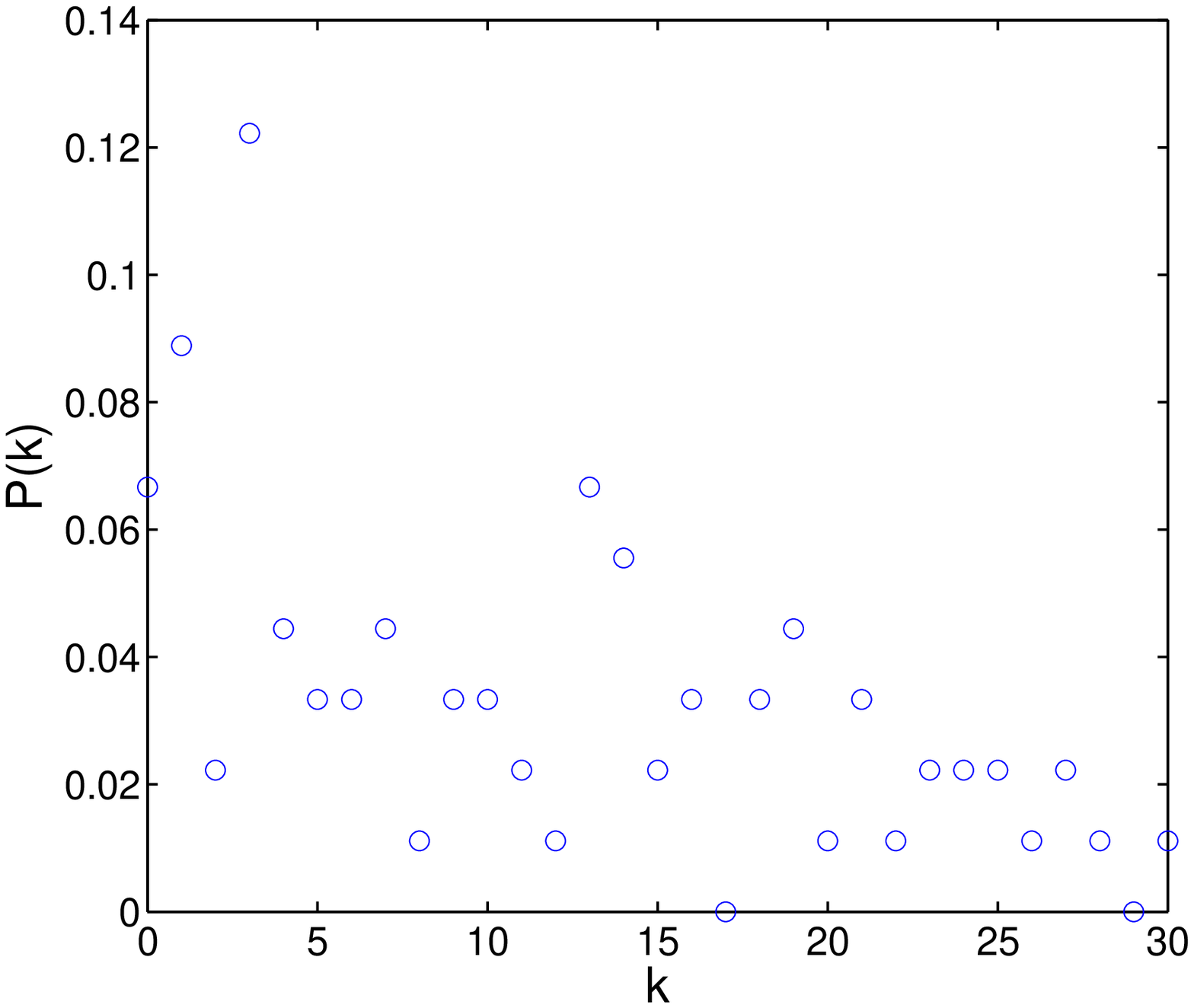}
\end{minipage}
\caption{Degree and degree distribution of the pathway-based gene 
network: (left) The values of the degree of $90$ nodes are shown in 
descending order; (right) The degree distribution is not scale-free.}
\label{degree and distribution}
\end{figure}

For the pathway-based gene network of hypertension, the degree 
$k_{i}$ of a node $i$ is just the number of other genes which are 
involved in the same pathway(s) as gene $i$. 
Fig.~\ref{degree and distribution} plots the degree and degree 
distribution of the gene network. Here, the values of the degree of 
$90$ nodes are ranked in a descending order (with $n_{g}(k)$ 
denoting the rank of genes in degree values), and it is found that 
the gene {\it Jun} of $n_{g}=1$ has the highest degree $30$. The average 
degree $\langle k \rangle$ of this network is about $10.71$. The 
illustration of degree distribution shows that the probability that 
a gene can link with $k$ other genes does not decay as a power-law, 
suggesting that the gene network does not have a scale-free topology.

\subsection{Average path length and clustering coefficient}
\label{subsec:3.2}

In a network, a path from node $i$ to node $j$ is a sequence of 
adjacent nodes starting at $i$ and ending at $j$. The path with the 
smallest number of edges between the two selected nodes is called the 
shortest path. The distance $d_{ij}$ between two nodes $i$ and $j$ 
is defined as the number of edges along the shortest path connecting 
them. The diameter $D$ is the maximum distance between any pair of 
nodes in the network, i.e., $D=\max\{d_{ij}\}$. The average path 
length $L$ is defined as the mean distance between two nodes, 
averaged over all pairs of nodes, i.e. 
\begin{eqnarray}
L=\frac{1}{N(N-1)}\sum_{i\neq j} d_{ij},
\end{eqnarray}
which offers a measure of the overall navigability of a 
network~\cite{X2003tag}.

A clustering coefficient can be used to describe the cohesiveness of 
the neighborhood of a node~\cite{WS1998n,N2003sr}. In a network, the 
clustering coefficient $c_{i}$ is defined as the ratio between the 
number $e_{i}$ of edges that actually link the $k_{i}$ neighbors of 
node $i$ to each other and the total possible number of edges among 
them, i.e. 
\begin{eqnarray}
c_{i}=\frac{2e_{i}}{k_{i}(k_{i}-1)} & (k_{i}\geqslant 2).
\label{clustering coef}
\end{eqnarray}
The clustering coefficient $C$ of the whole network is the average 
of $c_{i}$ over all $i$, $C=\frac{1}{N}\sum_{i=1}^{{N}} c_{i}$, 
which characterizes the overall tendency of nodes to form clusters, 
clearly, $C\leqslant 1$.

The pathway-based gene network has a very short average path length: 
$L$ is about $2.331$, and $\log(\log N)<L<\log N$. The diameter $D$ 
is only $5$ (i.e., the distance between genes {\it Gcgr} and 
{\it Gnpat}), which implies at most five hops separate any two genes 
in the $482$ connections of the network. The clustering coefficient 
is calculated to be $C=0.6403$, which is relatively high. Therefore, 
the pathway-based gene network has the small-world property (short 
$L$ and high $C$).

\subsection{Assortativity}
\label{subsec:3.3}

The concept of assortativity is introduced to describe degree 
correlations between neighboring nodes in a network~\cite{N2002prl}. 
A network is assortative if high (low) degree nodes tend to be 
connected to other high (low) degree nodes; otherwise, it is 
disassortative if high (low) degree nodes tend to be connected to 
other low (high) degree nodes. The assortativity can be described by 
the correlation between the degrees of neighboring nodes in terms of 
the mean Pearson correlation coefficient. Let $x_{i}$ and $y_{i}$ 
be the degrees of the end nodes of the $i$th edge, with 
$i=1,2,\ldots,E$ ($E$ is the number of edges in the network), then 
the assortativity coefficient of the network is given 
by~\cite{N2002prl}: 
\begin{eqnarray}
r=\frac{E^{-1}\sum\limits_{i}x_{i}y_{i}
  -\left[E^{-1}\sum\limits_{i}\frac{1}{2}(x_{i}+y_{i})\right]^{2}}
  {E^{-1}\sum\limits_{i}\frac{1}{2}(x_{i}^{2}+y_{i}^{2})
  -\left[E^{-1}\sum\limits_{i}\frac{1}{2}(x_{i}+y_{i})\right]^{2}}.
\label{assortativity coef}
\end{eqnarray}
The network is assortative if $r>0$, and disassortative if $r<0$. 
The assortativity coefficient of the pathway-based gene network is 
calculated to be $r=0.2178$, exhibiting an assortative behavior. This 
is different from most biological networks which show negative $r$.

\subsection{Centrality}
\label{subsec:3.4}

\subsubsection{Definitions of four centrality indices}
\label{subsubsec:3.4.1}

The degree centrality $C_{d}$, betweenness centrality $C_{b}$ and 
closeness centrality $C_{c}$ are three centrality indices commonly 
used in finding out the centralization nodes of the 
network~\cite{B1965bs,F1979sn}. Recently, we also introduced an 
integrated centrality $C_{\mathrm{intgr}}$ to fully reflect the 
contribution of three centrality indices 
$\{C_{d},C_{b},C_{c}\}$~\cite{WXHC2014pa}.

The degree centrality of a given node $i$ is the proportion of other 
nodes that are adjacent to node $i$~\cite{F1979sn}, i.e. 
\begin{eqnarray}
C_{d}(i)=\frac{k_{i}}{N-1},
\label{degree centrality}
\end{eqnarray}
here $N-1$ is the maximum possible degree of the network.

The betweenness centrality of a node $i$ is defined as the proportion 
of all shortest paths (geodesics) between pairs of other nodes that 
include this node $i$~\cite{F1979sn}: 
\begin{eqnarray}
C_{b}(i)=\sum_{j(<k)}^{N}\sum_{k}^{N}\frac{g_{jk}(i)}{g_{jk}},
\label{betweenness}
\end{eqnarray}
where $g_{jk}$ is the number of shortest paths between nodes $j$ and 
$k$, and $g_{jk}(i)$ the number of shortest paths containing node 
$i$ between nodes $j$ and $k$.

The closeness centrality of a node $i$ is the number of other nodes 
divided by the sum of the distances between node $i$ and all 
others~\cite{B1965bs,F1979sn}: 
\begin{eqnarray}
C_{c}(i)=(L_{i})^{-1}=\frac{N-1}{\sum\limits_{j=1}^{N}d_{ij}},
\label{closeness}
\end{eqnarray}
here $L_{i}$ is the average distance between node $i$ and all other 
nodes.

To comprehensively and quantitatively reflect the contribution of the 
above three centrality indices $\{C_{d},C_{b},C_{c}\}$, we can also 
introduce the integrated centrality $C_{\mathrm{intgr}}$ of node $i$, 
defined as follows~\cite{WXHC2014pa}: 
\begin{eqnarray}
C_{\mathrm{intgr}}(i)=\frac{1}{3}\left[\frac{C_{d}(i)}{C_{d,\max}}+%
\frac{C_{b}(i)}{C_{b,\max}}+\frac{C_{c}(i)}{C_{c,\max}}\right],
\label{intgr.centrality}
\end{eqnarray}
where $C_{d,\max}$, $C_{b,\max}$ and $C_{c,\max}$ are the maximums 
of $\{C_{d}\}$, $\{C_{b}\}$ and $\{C_{c}\}$, respectively. 
Obviously, $C_{\mathrm{intgr}}(i)$ has a value between $0$ and $1$.

\subsubsection{Centrality analysis and hub genes}
\label{subsubsec:3.4.2}

For the pathway-based gene network of hypertension, the four 
centrality indices are calculated based on the above definitions, 
and the top $25$ values of each centrality index are listed in 
Table~\ref{centrality data}. In the following, we will determine hub 
genes in the network through centrality analysis.

\begin{table}[t]
\centering
\scriptsize
\caption{Top $25$ values of degree centrality $C_{d}$, betweenness 
centrality $C_{b}$, closeness centrality $C_{c}$ and integrated 
centrality $C_{\mathrm{intgr}}$ in the pathway-based gene network of 
hypertension. In this paper, nine genes are expressed as 
abbreviations (cf. Fig.~\ref{dendrogram} and its caption).}
\label{centrality data}
\begin{tabular}[c]{llp{1mm}llp{1mm}llp{1mm}ll}
\hline
Gene & $C_{d}$   & &  Gene & $C_{b}$   & & Gene & $C_{c}$   & &  Gene & $C_{\mathrm{intgr}}$\\
\cline{1-2}\cline{4-5}\cline{7-8}\cline{10-11}
{\it Jun}      &  0.3371   & &  {\it Sdhb}     &  0.09327   & &  {\it Jun}      &  0.5087   
& &  {\it Jun}      &  0.9280\\
{\it Cdk4}     &  0.3146   & &  {\it Jun}      &  0.07311   & &  {\it Rps6kb1}  &  0.4880   
& &  {\it Rps6kb1}  &  0.8359\\
{\it RT1-Da}   &  0.3034   & &  {\it RT1-Da}   &  0.06903   & &  {\it Cycs}     &  0.4653   
& &  {\it Cycs}     &  0.8156\\
{\it Pdgfra}   &  0.3034   & &  {\it Cycs}     &  0.06057   & &  {\it Creb3l2}  &  0.4653   
& &  {\it Creb3l2}  &  0.8027\\
{\it Rps6kb1}  &  0.2921   & &  {\it Rps6kb1}  &  0.05791   & &  {\it Cdk4}     &  0.4446   
& &  {\it Cdk4}     &  0.7547\\
{\it Creb3l2}  &  0.2809   & &  {\it Cd36}     &  0.05748   & &  {\it Actg1}    &  0.4446   
& &  {\it Actg1}    &  0.7281\\
{\it Actg1}    &  0.2809   & &  {\it Pdha1}    &  0.04875   & &  {\it RT1-Da}   &  0.4413   
& &  {\it RT1-Da}   &  0.7128\\
{\it Csf1r}    &  0.2697   & &  {\it Creb3l2}  &  0.04067   & &  {\it Shc1}     &  0.4413   
& &  {\it Shc1}     &  0.6662\\
{\it Fn1}      &  0.2697   & &  {\it Fcgr1}    &  0.03694   & &  {\it Pdgfra}   &  0.4413   
& &  {\it Pdgfra}   &  0.6598\\
{\it Col4a1}   &  0.2584   & &  {\it Ctsl}     &  0.03632   & &  {\it Csf1r}    &  0.4381   
& &  {\it Csf1r}    &  0.6462\\
{\it Col4a2}   &  0.2584   & &  {\it Cdk4}     &  0.03088   & &  {\it Fn1}      &  0.4381   
& &  {\it Fn1}      &  0.6390\\
{\it Fzd2}     &  0.2472   & &  {\it Shc1}     &  0.02745   & &  {\it Cd36}     &  0.4318   
& &  {\it Cd36}     &  0.6293\\
{\it Cycs}     &  0.2360   & &  {\it Csf1r}    &  0.02586   & &  {\it Fzd2}     &  0.4287   
& &  {\it Fzd2}     &  0.6206\\
{\it Shc1}     &  0.2360   & &  {\it Acsl1}    &  0.02133   & &  {\it Col4a1}   &  0.4227   
& &  {\it Col4a1}   &  0.6041\\
{\it Tgfbr1}   &  0.2360   & &  {\it Acsl6}    &  0.02133   & &  {\it Col4a2}   &  0.4227   
& &  {\it Col4a2}   &  0.5956\\
{\it Fcgr1}    &  0.2247   & &  {\it Pdgfra}   &  0.01974   & &  {\it Tgfbr1}   &  0.4227   
& &  {\it Tgfbr1}   &  0.5841\\
{\it Col2a1}   &  0.2135   & &  {\it Actg1}    &  0.01955   & &  {\it Fcgr1}    &  0.4197   
& &  {\it Sdhb}     &  0.5678\\
{\it Col5a1}   &  0.2135   & &  {\it Ghr}      &  0.01847   & &  {\it Sdhb}     &  0.4111   
& &  {\it Fcgr1}    &  0.5678\\
{\it Col5a2}   &  0.2135   & &  {\it Nrp1}     &  0.01839   & &  {\it Mmp2}     &  0.4083   
& &  {\it Mmp2}     &  0.5640\\
{\it Col6a1}   &  0.2135   & &  {\it Ctss}     &  0.01467   & &  {\it Ctsl}     &  0.4056   
& &  {\it Ctsl}     &  0.5388\\
{\it Sdhb}     &  0.2022   & &  {\it Cdc2a}    &  0.01256   & &  {\it Col2a1}   &  0.4056   
& &  {\it Col2a1}   &  0.4980\\
{\it Mmp2}     &  0.2022   & &  {\it Fzd2}     &  0.01080   & &  {\it Col5a1}   &  0.4056   
& &  {\it Col5a1}   &  0.4980\\
{\it Ctsl}     &  0.2022   & &  {\it Polr3e}   &  0.01018   & &  {\it Col5a2}   &  0.4056   
& &  {\it Col5a2}   &  0.4980\\
{\it Cd36}     &  0.1798   & &  {\it Col4a1}   &  0.00987   & &  {\it Col6a1}   &  0.4056   
& &  {\it Col6a1}   &  0.4980\\
{\it Pdha1}    &  0.1798   & &  {\it Col4a2}   &  0.00987   & &  {\it Sdc1}     &  0.3949   
& &  {\it Sdc1}     &  0.4796\\
\hline
\end{tabular}
\end{table}

\begin{figure}[!]
\begin{center}
\includegraphics[width=15.1cm]{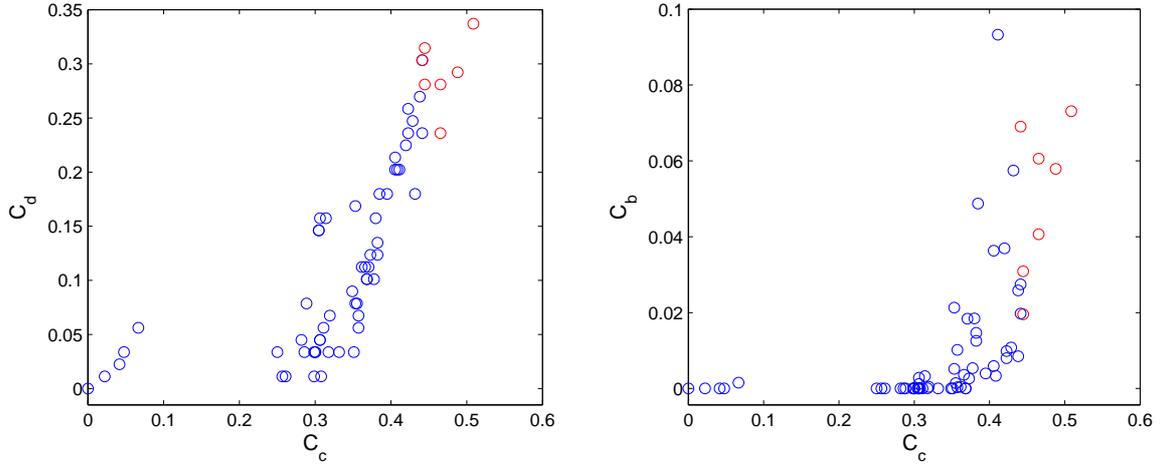}
\caption{Correspondence among degree centrality $C_{d}$, betweenness 
centrality $C_{b}$ and closeness centrality $C_{c}$ of nodes in the 
pathway-based gene network: (left) $C_{d}$ versus $C_{c}$; 
(right) $C_{b}$ versus $C_{c}$. The red circles represent the seven 
hub genes. 
(For interpretation of the references to color in this figure legend, 
the reader is referred to the web version of this article.)}
\label{CdCc and CbCc}
\end{center}
\end{figure}

\begin{figure}[!]
\begin{center}
\includegraphics[width=10.1cm]{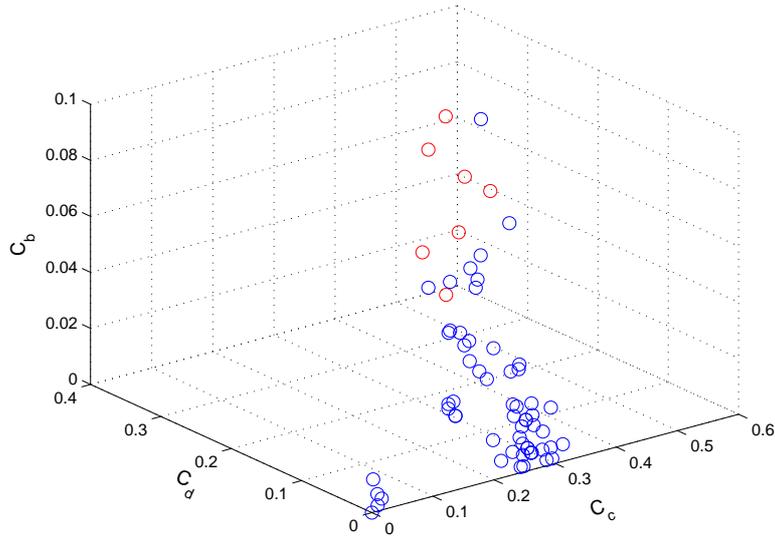}
\caption{Three-dimensional diagram of degree centrality $C_{d}$, 
betweenness centrality $C_{b}$ and closeness centrality $C_{c}$ of 
nodes in the pathway-based gene network. The red circles represent 
the seven hub genes. 
(For interpretation of the references to color in this figure legend, 
the reader is referred to the web version of this article.)}
\label{3D-CdCbCc}
\end{center}
\end{figure}

Fig.~\ref{CdCc and CbCc} describes the correspondence among degree 
centrality $C_{d}$, betweenness centrality $C_{b}$ and closeness 
centrality $C_{c}$ of nodes in the pathway-based gene network. We 
can see in Fig.~\ref{CdCc and CbCc} that $C_{c}$ of most of the 
nodes is distributed in the range of $(0.25,0.51)$. In this range of 
$C_{c}$, we also observe that in Fig.~\ref{CdCc and CbCc} (left) 
the distribution of $C_{d}$ is roughly uniform in the range of 
$(0.01,0.34)$; while in Fig.~\ref{CdCc and CbCc} (right) $C_{b}$ 
is non-uniformly distributed in the range of $[0,0.094)$, here 
most of the nodes have very small $C_{b}$, and only a few nodes 
have large $C_{b}$ which also have relatively large $C_{c}$. 
Fig.~\ref{3D-CdCbCc} combines the three centrality indices 
$\{C_{d},C_{b},C_{c}\}$ in the three-dimensional space. It can be 
seen from Fig.~\ref{3D-CdCbCc} that a small number of nodes have 
high values of three centrality indices, which can be viewed as hubs 
of the network. Generally speaking, each of these three centrality 
indices has its own focus on an influence of a node on other nodes 
in the network, thus one can identify hubs according to that focus. 
However, in order to fully reflect the contribution of all these 
three centrality indices, we will simply determine key hub genes 
using integrated centrality $C_{\mathrm{intgr}}$, as in 
Ref.~\cite{WXHC2014pa}.

\begin{figure}[!]
\begin{center}
\includegraphics[width=7.5cm]{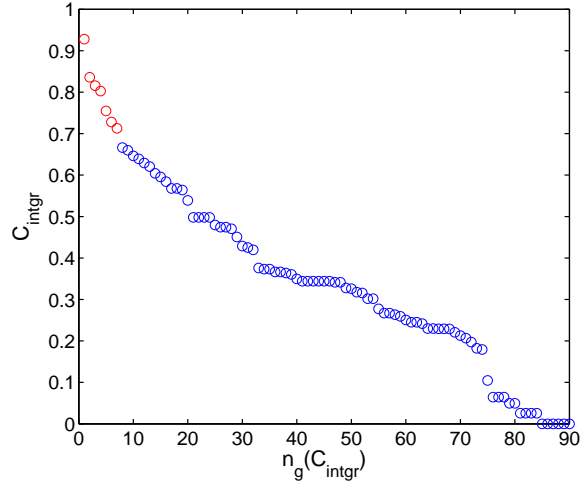}
\caption{The values of integrated centrality $C_{\mathrm{intgr}}$ of 
$90$ nodes in descending order for the pathway-based gene network. 
The red circles represent the seven hub genes. 
(For interpretation of the references to color in this figure legend, 
the reader is referred to the web version of this article.)}
\label{Cintgr.vs.ng}
\end{center}
\end{figure}

As calculated and shown above, the pathway-based gene network is not 
scale-free. Fig.~\ref{Cintgr.vs.ng} illustrates the integrated 
centrality $C_{\mathrm{intgr}}$ of $90$ nodes in a descending order 
(with $n_{g}(C_{\mathrm{intgr}})$ denoting the rank of genes in 
$C_{\mathrm{intgr}}$) for the network, which shows that the top seven 
nodes (red) have high values of $C_{\mathrm{intgr}}$. From 
Table~\ref{centrality data} and Fig.~\ref{Cintgr.vs.ng}, seven nodes 
are determined as important hub genes: 
{\it Jun}, {\it Rps6kb1}, {\it Cycs}, {\it Creb3l2}, {\it Cdk4}, 
{\it Actg1} and {\it RT1-Da}. 
These seven genes have the integrated centrality of 
$C_{\mathrm{intgr}}>0.71$, which is about $77\%$ of its maximum 
($0.9280$). However, it should be mentioned that there is no strict 
significance threshold for $C_{\mathrm{intgr}}$, and one can also lower 
the threshold to enable more genes to be included in hub genes.

In Table~\ref{hubs descriptions} we list the official full names, gene 
IDs and biological descriptions of seven hub genes. Here, we also 
provide a brief biological description of the first hub gene. The 
gene {\it Jun} is ranked first in $C_{\mathrm{intgr}}$ because it has the 
highest $C_{d}$ and $C_{c}$, as well as the second highest $C_{b}$. 
{\it Jun} encodes a protein that exhibits double-stranded DNA binding, 
involved in aging, angiogenesis, endothelin and Rho/Rac/Cdc42 
mediated signaling pathways. It is associated with kidney neoplasms 
and spinal cord injuries. Recent studies have confirmed that {\it Jun} is 
closely related to low potassium reaction and cell renal cell 
carcinoma. A low potassium diet might induce hypertension, which is 
always accompanied by hypokalemia. The incidence of renal cell 
carcinoma coupled with hypertension is up to 
$14\%$--$40\%$~\cite{KKSS2012jccr}.

\begin{table}[!t]
\caption{Biological descriptions of seven hub genes in the 
pathway-based gene network.}
\label{hubs descriptions}
\centering
\scriptsize
\def\temptablewidth{1\textwidth}
\begin{tabular*}{\temptablewidth}{@{\extracolsep{\fill}}lp{5.2cm}lp{6.3cm}}
\hline
Gene    & Official full name & Gene ID         & Description\\
\hline 
{\it Jun}     & jun proto-oncogene & {\;\,}24{\,}516 & 
Encodes a protein that exhibits double-stranded DNA binding\\
{\it Rps6kb1} & ribosomal protein S6 kinase, polypeptide 1 & {\;\,}83{\,}840 & 
Encodes a protein that exhibits ATP binding; peptide binding\\
{\it Cycs}    & cytochrome c, somatic & {\;\,}25{\,}309 & 
Encodes a protein that exhibits electron carrier activity\\
{\it Creb3l2} & cAMP responsive element binding protein 3-like 2 & 362{\,}339 & 
Encodes a protein that exhibits cAMP response element binding\\
{\it Cdk4}    & cyclin-dependent kinase 4 & {\;\,}94{\,}201 & 
Related to circadian rhythm; organ regeneration\\
{\it Actg1}   & actin, gamma 1 & 287{\,}876 & 
Involved in response to calcium ion\\
{\it RT1-Da}  & RT1 class \Rmnum{2}, locus Da & 294{\,}269 & 
Involved in antigen processing\\
\hline
\end{tabular*}
\end{table}

Among the seven hub genes, we note that the gene {\it Cdk4} is the 
only common hub gene in both the pathway-based gene network here and 
the gene co-expression network of our previous 
study~\cite{WXHC2014pa}. Besides {\it Cdk4}, we also see in 
Table~\ref{centrality data} that the three genes 
{\it Shc1}, {\it Fzd2} and {\it Col4a1}, which have relatively high 
integrated centrality $C_{\mathrm{intgr}}>0.60$, are the hub genes 
identified in the gene co-expression network of 
Ref.~\cite{WXHC2014pa}. These four genes have been confirmed by 
biological and medical research to play important roles in 
hypertension. Moreover, we also observe that although the gene 
{\it Sdhb} is ranked only joint 17th in $C_{\mathrm{intgr}}$ 
($=0.5678$), it has the highest betweenness centrality $C_{b}$. Since 
$C_{b}$ is based on the shortest paths and reflects the ability of a 
node to influence other related nodes in the network, {\it Sdhb} 
should also be a key gene in hypertension. If we lower the threshold 
of integrated centrality to $C_{\mathrm{intgr}}>0.50$, then these four 
genes ({\it Shc1}, {\it Fzd2}, {\it Col4a1} and {\it Sdhb}) can also 
be included in hub genes. In this paper, we do not take this lower 
threshold of $C_{\mathrm{intgr}}>0.50$ because, in view of the 
relatively large range of variation of $C_{b}$, we do not want the 
genes of small $C_{b}$ to be included in hub genes. In a wider view, 
however, these four genes ({\it Shc1}, {\it Fzd2}, {\it Col4a1} and 
{\it Sdhb}), together with the above seven hub genes, are worthy of 
further study in the future.

\section{Modular structure of the gene network}
\label{sec:4}

Many networks are found to divide naturally into modules or 
communities, i.e., groups of nodes within which the connections are 
relatively dense but between which they are 
sparser~\cite{GN2002pnas,N2006pnas}. In this section, we explore the 
modular structure of the pathway-based gene network of hypertension.

\subsection{Structural equivalence of nodes}
\label{subsec:4.1}

Two nodes are structural equivalent if they have identical 
connections with all other nodes. We can use a dissimilarity index 
$d_{s}$ to measure the equivalence of two nodes $i$ and $j$ as 
follows~\cite{dNMB2005esnap}: 
\begin{eqnarray}
d_{s}(i,j)=\frac{|V(i)+V(j)|}{k_{1}+k_{2}}.
\label{dissimilarity index}
\end{eqnarray}
Here $i,j=1,2,\ldots,90$, $V(i)$ are all neighbors of node 
$i$, $|\cdot |$ stands for set cardinality, $k_{1}$ and $k_{2}$ 
stand for the largest and the second largest degree in the network, 
respectively. Obviously, $d_{s}(i,j)$ has a value between $0$ 
(completely similar) and $1$ (completely different). In 
Table~\ref{dissimilarity scores} we list the dissimilarity scores of ten 
genes distributed in different modules (cf. Fig.~\ref{modules}).

\begin{table}[!tbp]
\caption{Dissimilarity scores $d_{s}(i,j)$ of ten nodes (genes) in 
the pathway-based gene network. The specific modules, which are 
identified after the completion of the modular decomposition, are 
indicated in parentheses in the first column.}
\label{dissimilarity scores}
\centering
\scriptsize
\begin{tabular}[c]{lllllllllll} 
\hline
                     & {\it Jun} & {\it RT1-Da} & {\it Col4a1} & {\it Fzd2} & {\it Sdhb} & {\it Gda} & {\it Sec13l1} 
                     & {\it Sumo1} & {\it Aqp1} & {\it Kcnj1}\\
\hline
{\it Jun} (\Rmnum{1})     & 0.0000 & 0.4310 & 0.2931 & 0.2414 & 0.7241 & 0.6897 & 0.6034 & 0.5517 & 0.5345 & 0.5172\\
{\it RT1-Da} (\Rmnum{1})  & 0.4310 & 0.0000 & 0.5172 & 0.3621 & 0.7069 & 0.7069 & 0.5517 & 0.5000 & 0.4828 & 0.4655\\
{\it Col4a1} (\Rmnum{1})  & 0.2931 & 0.5172 & 0.0000 & 0.2931 & 0.6034 & 0.6379 & 0.4828 & 0.4310 & 0.4138 & 0.3966\\
{\it Fzd2} (\Rmnum{1})    & 0.2414 & 0.3621 & 0.2931 & 0.0000 & 0.5862 & 0.6207 & 0.4655 & 0.4138 & 0.3966 & 0.3793\\
{\it Sdhb} (\Rmnum{2})    & 0.7241 & 0.7069 & 0.6034 & 0.5862 & 0.0000 & 0.1034 & 0.3966 & 0.3448 & 0.3276 & 0.3103\\
{\it Gda} (\Rmnum{2})     & 0.6897 & 0.7069 & 0.6379 & 0.6207 & 0.1034 & 0.0000 & 0.3276 & 0.2759 & 0.2586 & 0.2414\\
{\it Sec13l1} (\Rmnum{3}) & 0.6034 & 0.5517 & 0.4828 & 0.4655 & 0.3966 & 0.3276 & 0.0000 & 0.0517 & 0.1034 & 0.0862\\
{\it Sumo1} (\Rmnum{3})   & 0.5517 & 0.5000 & 0.4310 & 0.4138 & 0.3448 & 0.2759 & 0.0517 & 0.0000 & 0.0517 & 0.0345\\
{\it Aqp1} (\Rmnum{4})    & 0.5345 & 0.4828 & 0.4138 & 0.3966 & 0.3276 & 0.2586 & 0.1034 & 0.0517 & 0.0000 & 0.0172\\
{\it Kcnj1} (\Rmnum{5})   & 0.5172 & 0.4655 & 0.3966 & 0.3793 & 0.3103 & 0.2414 & 0.0862 & 0.0345 & 0.0172 & 0.0000\\
\hline
\end{tabular}
\end{table}

\begin{figure}[p]
\begin{center}
\includegraphics[width=11.4cm]{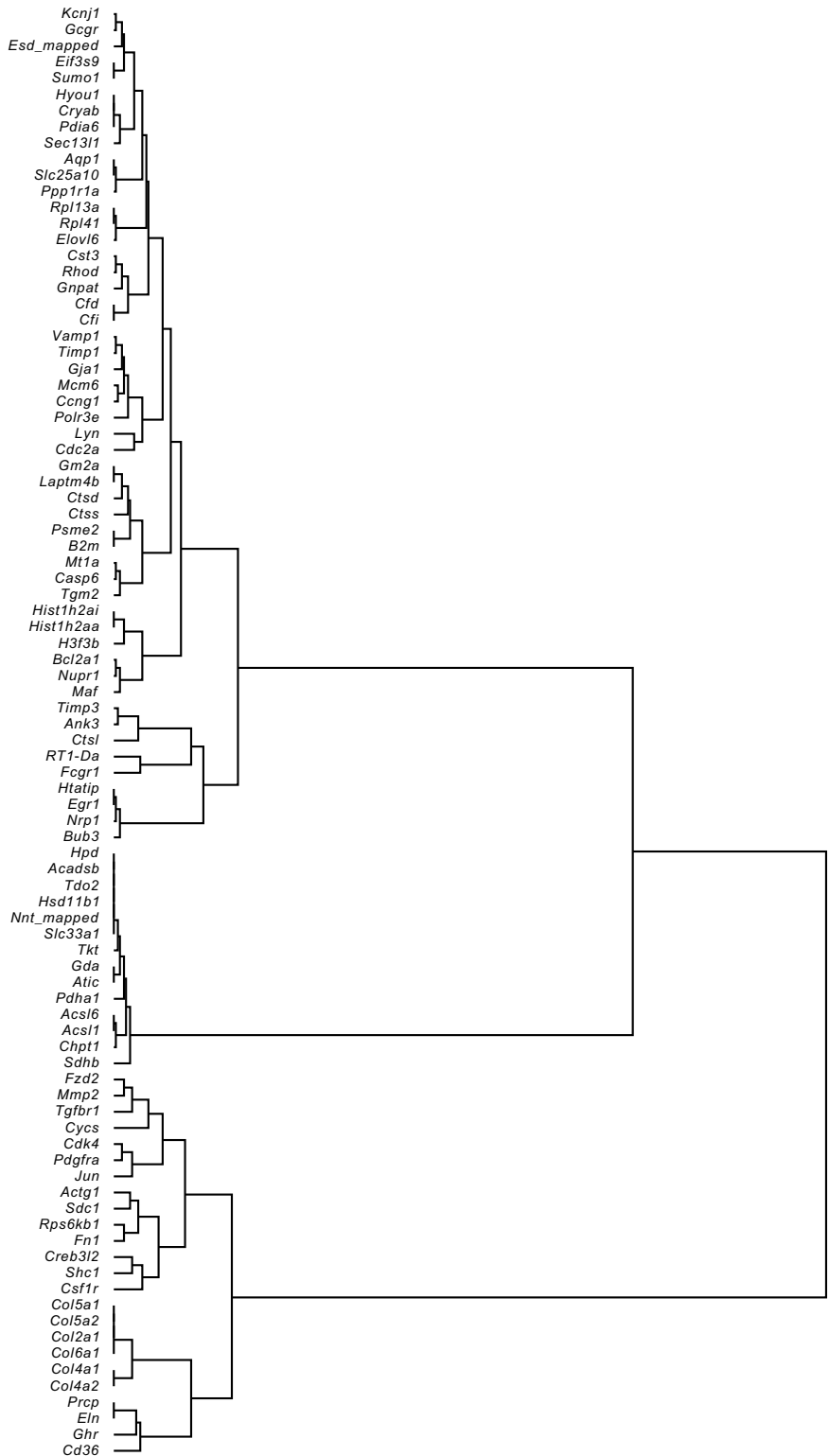}
\caption{Dendrogram of similarities. It shows the hierarchical 
clustering of the pathway-based gene network. In this paper, the 
following nine genes (as recorded in Ref.~\cite{LLW2008pg}), 
{\it Rhod\_predicted}, 
{\it Polr3e\_predicted}, 
{\it Hist1h2ai\_predicted /// Hist1h4a\_predicted}, 
{\it Sdhb\_predicted}, 
{\it Actg1 /// LOC295810}, 
{\it Col6a1\_predicted}, 
{\it Col4a2\_predicted}, 
{\it Prcp\_predicted}, 
and 
{\it Cd36 /// RGD1562323\_predicted}, 
are abbreviated as 
{\it Rhod}, {\it Polr3e}, {\it Hist1h2ai}, {\it Sdhb}, {\it Actg1}, 
{\it Col6a1}, {\it Col4a2}, {\it Prcp}, and {\it Cd36}, 
respectively.}
\label{dendrogram}
\end{center}
\end{figure}

The dissimilarity scores allow us to cluster nodes in accordance with 
the structural equivalence into the corresponding positions by the 
hierarchical clustering technique. First, the nodes that are most 
similar are grouped into a cluster. Then, the next pair of nodes or 
clusters that are most similar are grouped, and this process continues 
until all nodes have been joined. The dendrogram in 
Fig.~\ref{dendrogram} is obtained with Pajek 
software~\cite{dNMB2005esnap,BM1998c}, which visualizes the above 
clustering process.

\subsection{Modular decomposition of the network}
\label{subsec:4.2}

\begin{figure}[!]
\begin{center}
\includegraphics[width=10cm]{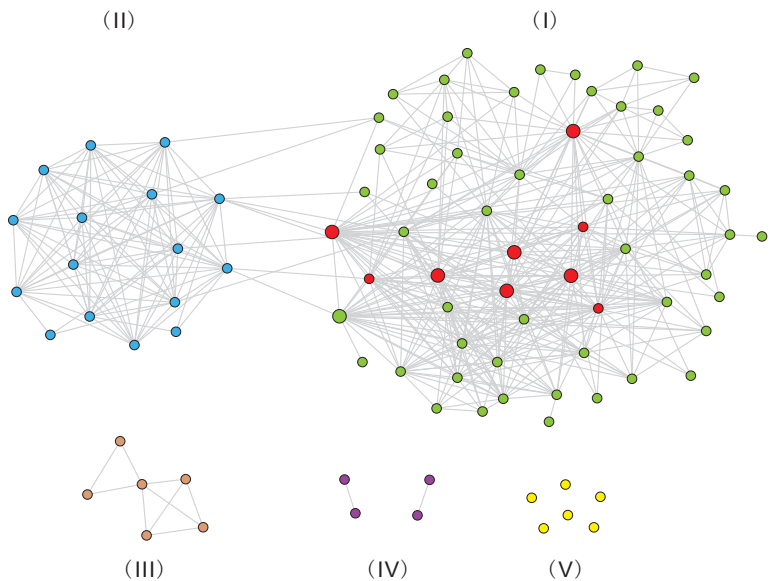}
\caption{Modular structure of the pathway-based gene network. Modules 
\Rmnum{1}--\Rmnum{5} contain $58$, $16$, $6$, $4$ and $6$ nodes, 
respectively. The nine nodes marked red in module \Rmnum{1} 
correspond to the top nine genes involved in the largest number of 
pathways. The seven large nodes (six red and one green) indicate the 
hub genes. 
(For interpretation of the references to color in this figure legend, 
the reader is referred to the web version of this article.)}
\label{modules}
\end{center}
\end{figure}

Based on the above dendrogram, we can obtain the modular structure of 
the gene network, which is shown in Fig.~\ref{modules}. The network 
consists of five modules: 
\begin{list}{}{}
\item (\Rmnum{1}) the largest module with $58$ nodes (red and green);
\item (\Rmnum{2}) the second largest module with $16$ nodes (blue);
\item (\Rmnum{3}) a small module with the highest clustering 
coefficient including $6$ nodes (brown);
\item (\Rmnum{4}) two pairs of adjacent nodes (each joined by a 
single connection) (purple);
\item (\Rmnum{5}) six isolated nodes (yellow).
\end{list}
We can observe that modules \Rmnum{1} and \Rmnum{2} constitute the 
largest connected part of the network, containing $74$ nodes and 
$471$ edges; and the nodes are highly connected within a module but 
much less connected between modules.

\begin{table}[!h]
\caption{Statistical characteristics of five modules in the 
pathway-based gene network.}
\label{stat.modules}
\centering
\scriptsize
\begin{tabular}[c]{llrlrllll}
\hline
Module     &&  $N_{m}$  && $N_{m}/N${\ \ \,} && $\langle k\rangle_{m}$ &&  $C_{m}$\\
\hline
\Rmnum{1}  &&   58{\ }  &&    64.44\%        &&       12.72              &&  0.6545\\
\Rmnum{2}  &&   16{\ }  &&    17.78\%        &&       12.75              &&  0.8917\\
\Rmnum{3}  &&    6{\ }  &&     6.67\%        &&  {\ \ }3                 &&  0.9000\\
\Rmnum{4}  &&    4{\ }  &&     4.44\%        &&  {\ \ }1                 &&  0\\
\Rmnum{5}  &&    6{\ }  &&     6.67\%        &&  {\ \ }0                 &&  0\\
\hline
\end{tabular}
\end{table}

In Table~\ref{stat.modules} we list several statistical 
characteristics of every module, including the number $N_{m}$ of 
nodes, node proportion $N_{m}/N$, average degree 
$\langle k\rangle_{m}$ and clustering coefficient $C_{m}$ 
($m=\mathrm{\Rmnum{1}},\mathrm{\Rmnum{2}},\mathrm{\Rmnum{3}},\mathrm{\Rmnum{4}},\mathrm{\Rmnum{5}}$). 
Here $C_{m}$ is the average of $c_{i}$ over all nodes in module $m$. 
From Table~\ref{stat.modules} we can see that the average degree of 
each of the two modules \Rmnum{1} and \Rmnum{2} 
($\langle k\rangle_{\mathrm{\Rmnum{1}}}=12.72$, 
$\langle k\rangle_{\mathrm{\Rmnum{2}}}=12.75$) is greater than the 
average degree $\langle k \rangle=10.71$ of the whole network. 
Except modules \Rmnum{4} and \Rmnum{5} of 
$C_{\mathrm{\Rmnum{4}}}=C_{\mathrm{\Rmnum{5}}}=0$, the clustering 
coefficient of each of other three modules (\Rmnum{1}--\Rmnum{3}) 
exceeds that of the whole network ($C=0.6403$), showing that there 
are more connections within each of these three modules, which 
justifies the modular decomposition of the whole network.

\subsection{Characteristics of modular structure}
\label{subsec:4.3}

The connections between nodes in the gene network are created based 
on whether genes are involved in the same pathway(s). Having examined 
the number of pathways in which each gene is involved, we find that 
there are nine genes involved in more than 10 pathways: 
{\it Jun} ($30$), {\it RT1-Da} ($21$), {\it Cdk4} ($17$), 
{\it Cycs} ($17$), {\it Creb3l2} ($16$), {\it Actg1} ($15$), 
{\it Pdgfra} ($14$), {\it Tgfbr1} ($14$) and {\it Shc1} ($13$), 
here the number of pathways involved is shown in parentheses. These 
nine genes (red nodes in Fig.~\ref{modules}) are a basis of dense 
connections within the largest module \Rmnum{1}. We also note that 
among these nine genes, there are six hub genes (large nodes marked 
red); another hub gene {\it Rps6kb1} (large node marked green) is 
involved in nine pathways.

We can examine the robustness of the network based on modular 
structure~\cite{N2006pnas,CNSW2000prl}. The nodes within a module 
(except \Rmnum{4} and \Rmnum{5} in this paper) are relatively robust 
against mutation because there are multiple paths between any two 
nodes and thus the network will not be easily broken when mutation 
occurs. Nevertheless, the parts of fewer connections between modules 
should be the weaknesses of the network system. Fig.~\ref{modules} 
visualizes the weak links of the network. The removal of these weak 
connections and relevant nodes would result in the breaking of the 
network. Thus, the modular structure analysis can facilitate the 
exploration of the relationship between weak connections of the 
network and drug targets of hypertension.

The genes in the largest connected part (i.e., core modules \Rmnum{1} 
and \Rmnum{2}) are involved in multiple pathways, and contribute to 
a variety of biological functions. It is difficult to assign each of 
these genes to a single biological function because of pleiotropy, 
namely, one gene might influence many different biological processes 
in organisms~\cite{HZ2012g}. However, the genes in the small non-core 
module \Rmnum{3} with six nodes are involved in only two specific 
pathways, rno03013 (RNA transport) and rno04141 (protein processing 
in endoplasmic reticulum (ER)), which indicates that the non-core 
module \Rmnum{3} corresponds to clearly identified pathways and 
functions relatively independently. 

In the pathway rno03013, the different RNA species produced in the 
nucleus are exported through the nuclear pore complexes (NPCs) to the 
cytoplasm via mobile export receptors, which is fundamental for gene 
expression. In the pathway rno04141, newly synthesized peptides enter 
the ER via the sec61 pore and are glycosylated; correctly folded 
proteins are packaged into transport vesicles and misfolded proteins 
are retained within the ER lumen (cf. KEGG PATHWAY Database).

\section{Summary and concluding remarks}
\label{sec:5}

Hypertension is a cardiovascular disease associated with long-term 
interaction between genetic and environmental factors. In this study, 
we use pathways data to obtain backwards the relationships between 
the hypertension-related genes, try to extract the complex 
interactions between genes through calculating statistical 
characteristics and analyzing modular structure of the network.

The pathway-based gene network has the following characteristics: 
(\rmnum{1}) The network does not obey a power-law degree distribution 
and thus is not of a scale-free property. The seven hub genes that 
are identified by integrated centrality $C_{\mathrm{intgr}}>0.71$ are: 
{\it Jun}, {\it Rps6kb1}, {\it Cycs}, {\it Creb3l2}, {\it Cdk4}, 
{\it Actg1} and {\it RT1-Da}; they are key (feature) genes involved 
in the formation of hypertension. 
(\rmnum{2}) The network shows the small-world property (i.e., a small 
$L$ and a large $C$), which reveals the direct influence of these hub 
genes on hypertension from another perspective. 
(\rmnum{3}) The network has a modular structure. The weak connections 
of the network can be visualized by its modular structure, which can 
help to screen out key hypertension-related genes or pathways.

In this paper, we construct the network model of hypertension-related 
genes based on biological pathways. Among the seven hub genes 
identified in this network, only {\it Cdk4} is also a hub gene in the 
gene co-expression network of our previous study~\cite{WXHC2014pa}. 
Besides {\it Cdk4}, the three genes {\it Shc1}, {\it Fzd2} and 
{\it Col4a1} with $C_{\mathrm{intgr}}>0.60$ in the pathway-based gene 
network are identified as the hub genes in the gene co-expression 
network of Ref.~\cite{WXHC2014pa}. Although we can see that more 
nodes will become hub genes and thus there will be more hub genes 
overlapped in the both networks if we lower the threshold of 
$C_{\mathrm{intgr}}$, the hub genes in the two networks are impossible 
to be completely overlapped because the two networks are constructed 
from the different perspectives, i.e., based on the different 
characters of the genes. The results from this paper and the 
theoretical analysis in Ref.~\cite{WXHC2014pa} would complement each 
other. The seven hub genes in the pathway-based gene network, together 
with the above-mentioned {\it Shc1}, {\it Fzd2} and {\it Col4a1}, 
as well as {\it Sdhb} of the highest $C_{b}$, can be regarded as 
candidate genes or drug targets for further biological and medical 
research on their functions; in particular, the common hub gene 
{\it Cdk4} of the both networks would be worth more attention.

Moreover, the network may also be analyzed based on other functional 
correlation methods, such as GO analysis~\cite{CGBC2011itbe}, to get 
more molecular mechanisms about hypertension. In the next study, we 
will develop weighted network models and explore the mutual 
regulatory relationships between genes of complex diseases using 
dynamical analysis. These studies will provide another perspective on 
expounding the differentially expressed genes and finding new drug 
targets for other serious diseases. Finally, we expect that the 
complex network approach can provide clues for exploring the 
pathogenesis of critical illness from molecular perspective.

\section*{Acknowledgments}

This work was supported in part by the National Natural Science 
Foundation of China (NSFC) (Grant Nos. 11365023 and 61263043), 
the Projects of the Science and Technology Research and Development 
Program of Baoji City (Grant Nos. 15RKX-1-5-14 and 15RKX-1-5-6), 
the Key Project of Baoji University of Arts and Sciences (Grant No. 
ZK14035), and the Joint Fund of Department of Science and Technology 
of Guizhou Province, Bureau of Science and Technology of Qiandongnan 
Prefecture, and Kaili University (Grant No. LH-2014-7231). We are 
grateful to the authors of Ref.~\cite{LLW2008pg} for providing the 
gene information of the SS rat, which is the basis of construction 
of our network model. The authors would like to thank Professor 
Huai Cao for his helpful discussions and suggestions.












\begin{thebibliography}{99}

\bibitem{WS1998n}
D.J. Watts, S.H. Strogatz, 
Collective dynamics of `small-world' networks, 
Nature 393 (6684) (1998) 440--442.

\bibitem{BA1999s}
A.-L. Barab\'{a}si, R. Albert, 
Emergence of scaling in random networks, 
Science 286 (5439) (1999) 509--512.

\bibitem{S2001n}
S.H. Strogatz, 
Exploring complex networks, 
Nature 410 (6825) (2001) 268--276.

\bibitem{N2003sr}
M.E.J. Newman, 
The structure and function of complex networks, 
SIAM Rev. 45 (2) (2003) 167--256.

\bibitem{WYY2006pre}
W.-X. Wang, C.-Y. Yin, G. Yan, B.-H. Wang, 
Integrating local static and dynamic information for routing traffic, 
Phys. Rev. E 74 (1) (2006) 016101.

\bibitem{JTA2000n}
H. Jeong, B. Tombor, R. Albert, Z.N. Oltvai, A.-L. Barab\'{a}si, 
The large-scale organization of metabolic networks, 
Nature 407 (6804) (2000) 651--654.

\bibitem{BO2004nrg}
A.-L. Barab\'{a}si, Z.N. Oltvai, 
Network biology: Understanding the cell's functional organization, 
Nature Rev. Genet. 5 (2) (2004) 101--113.

\bibitem{KH2007pa}
F. Karlsson, M. H\"{o}rnquist, 
Order or chaos in Boolean gene networks depends on the mean fraction of canalizing functions, 
Physica A 384 (2) (2007) 747--757.

\bibitem{TSP2009pa}
M. Tsuchiya, K. Selvarajoo, V. Piras, M. Tomita, A. Giuliani, 
Local and global responses in complex gene regulation networks, 
Physica A 388 (8) (2009) 1738--1746.

\bibitem{D2011pa}
L. Diambra, 
Coarse-grain reconstruction of genetic networks from expression levels, 
Physica A 390 (11) (2011) 2198--2207.

\bibitem{CO2000c}
O.A. Carretero, S. Oparil, 
Essential hypertension: Part I: Definition and etiology, 
Circulation 101 (3) (2000) 329--335.

\bibitem{CNB2012h}
A.W. Cowley Jr., J.H. Nadeau, A. Baccarelli, K. Berecek, M. Fornage, 
G.H. Gibbons, D.G. Harrison, M. Liang, P.W. Nathanielsz, D.T. O'Connor, 
J. Ordovas, W. Peng, M.B. Soares, M. Szyf, H.E. Tolunay, 
K.C. Wood, K. Zhao, Z.S. Galis, 
Report of the National Heart, Lung, and Blood Institute Working Group on epigenetics and hypertension, 
Hypertension 59 (5) (2012) 899--905.

\bibitem{WFF2001h}
M.H. Weinberger, N.S. Fineberg, S.E. Fineberg, M. Weinberger, 
Salt sensitivity, pulse pressure, and death in normal and hypertensive humans, 
Hypertension 37 (2) (2001) 429--432.

\bibitem{A2002ije}
M.H. Alderman, 
Salt, blood pressure and health: A cautionary tale, 
Int. J. Epidemiol. 31 (2) (2002) 311--315.

\bibitem{RIV2007ndt}
B. Rodriguez-Iturbe, N.D. Vaziri, 
Salt-sensitive hypertension---update on novel findings, 
Nephrol. Dial. Transplant. 22 (4) (2007) 992--995.

\bibitem{R2000pr}
J.P. Rapp, 
Genetic analysis of inherited hypertension in the rat, 
Physiol. Rev. 80 (1) (2000) 135--172.

\bibitem{C2006nrg}
A.W. Cowley Jr., 
The genetic dissection of essential hypertension, 
Nature Rev. Genet. 7 (11) (2006) 829--840.

\bibitem{CCDN2000c}
R. Cooper, J. Cutler, P. Desvigne-Nickens, S.P. Fortmann, L. Friedman, 
R. Havlik, G. Hogelin, J. Marler, P. McGovern, G. Morosco, 
L. Mosca, T. Pearson, J. Stamler, D. Stryer, T. Thom, 
Trends and disparities in coronary heart disease, stroke, and other 
cardiovascular diseases in the United States: Findings of the 
National Conference on Cardiovascular Disease Prevention, 
Circulation 102 (25) (2000) 3137--3147.

\bibitem{WMCK2004h}
K. Wolf-Maier, R.S. Cooper, H. Kramer, J.R. Banegas, S. Giampaoli, 
M.R. Joffres, N. Poulter, P. Primatesta, B. Stegmayr, M. Thamm, 
Hypertension treatment and control in five European countries, Canada, and the United States, 
Hypertension 43 (1) (2004) 10--17.

\bibitem{SJJ2011chr}
H. Sanada, J.E. Jones, P.A. Jose, 
Genetics of salt-sensitive hypertension, 
Curr. Hypertens. Rep. 13 (1) (2011) 55--66.

\bibitem{CGBC2011itbe}
F. Censi, A. Giuliani, P. Bartolini, G. Calcagnini, 
A multiscale graph theoretical approach to gene regulation networks: 
A case study in atrial fibrillation, 
IEEE Trans. Biomed. Eng. 58 (10) (2011) 2943--2946.

\bibitem{DC2011ao}
R. Demicheli, D. Coradini, 
Gene regulatory networks: A new conceptual framework to analyse breast cancer behaviour, 
Ann. Oncol. 22 (6) (2011) 1259--1265.

\bibitem{WXHC2014pa}
H. Wang, C.-Y. Xu, J.-B. Hu, K.-F. Cao, 
A complex network analysis of hypertension-related genes, 
Physica A 394 (2014) 166--176.

\bibitem{LLW2008pg}
M. Liang, N.H. Lee, H. Wang, A.S. Greene, A.E. Kwitek, 
M.L. Kaldunski, T.V. Luu, B.C. Frank, S. Bugenhagen, H.J. Jacob, 
A.W. Cowley Jr., 
Molecular networks in Dahl salt-sensitive hypertension based on 
transcriptome analysis of a panel of consomic rats, 
Physiol. Genomics 34 (1) (2008) 54--64.

\bibitem{DHT1962jem}
L.K. Dahl, M. Heine, L. Tassinari, 
Effects of chronic excess salt ingestion: Evidence that genetic factors 
play an important role in susceptibility to experimental hypertension, 
J. Exp. Med. 115 (6) (1962) 1173--1190.

\bibitem{R1982h}
J.P. Rapp, 
Dahl salt-susceptible and salt-resistant rats: A review, 
Hypertension 4 (6) (1982) 753--763.

\bibitem{D1998jh}
A.Y. Deng, 
In search of hypertension genes in Dahl salt-sensitive rats, 
J. Hypertens. 16 (12) (1998) 1707--1717.

\bibitem{CRJ2004jp}
A.W. Cowley Jr., R.J. Roman, H.J. Jacob, 
Application of chromosomal substitution techniques in gene-function discovery, 
J. Physiol. 554 (1) (2004) 46--55.

\bibitem{K2002s}
H. Kitano, 
Systems biology: A brief overview, 
Science 295 (5560) (2002) 1662--1664.

\bibitem{DGvH2003bb}
Y. Deville, D. Gilbert, J. van Helden, S.J. Wodak, 
An overview of data models for the analysis of biochemical pathways, 
Brief. Bioinform. 4 (3) (2003) 246--259.

\bibitem{JLG2005pg}
B. Joe, N.E. Letwin, M.R. Garrett, S. Dhindaw, B. Frank, 
R. Sultana, K. Verratti, J.P. Rapp, N.H. Lee, 
Transcriptional profiling with a blood pressure QTL interval-specific oligonucleotide array, 
Physiol. Genomics 23 (3) (2005) 318--326.

\bibitem{W2013DScD}
H. Wang, 
Analyses of hypertension-related genes based on complex network theory 
(Doctor of Science Dissertation), Yunnan University, Kunming, 2013 
(in Chinese, with English abstract).

\bibitem{X2003tag}
J. Xu, 
Theory and Application of Graphs, 
Kluwer Academic Publishers, Dordrecht, Boston, London, 2003.

\bibitem{N2002prl}
M.E.J. Newman, 
Assortative mixing in networks, 
Phys. Rev. Lett. 89 (20) (2002) 208701.

\bibitem{B1965bs}
M.A. Beauchamp, 
An improved index of centrality, 
Behav. Sci. 10 (2) (1965) 161--163.

\bibitem{F1979sn}
L.C. Freeman, 
Centrality in social networks: Conceptual clarification, 
Social Networks 1 (3) (1978--1979) 215--239.

\bibitem{KKSS2012jccr}
E. Konik, E.G. Kurtz, F. Sam, D. Sawyer, 
Coronary artery spasm, hypertension, hypokalemia and licorice, 
J. Clin. Case Rep. 2 (8) (2012) 143.

\bibitem{GN2002pnas}
M. Girvan, M.E.J. Newman, 
Community structure in social and biological networks, 
Proc. Natl. Acad. Sci. USA 
99 (12) (2002) 7821--7826.

\bibitem{N2006pnas}
M.E.J. Newman, 
Modularity and community structure in networks, 
Proc. Natl. Acad. Sci. USA 
103 (23) (2006) 8577--8582.

\bibitem{dNMB2005esnap}
W. de Nooy, A. Mrvar, V. Batagelj, 
Exploratory Social Network Analysis with Pajek, 
in: Structural Analysis in the Social Sciences, vol. 34, 
Cambridge University Press, New York, 2005; 
Revised and expanded second edition, 2011.

\bibitem{BM1998c}
V. Batagelj, A. Mrvar, 
Pajek: A program for large network analysis, 
Connections 21 (2) (1998) 47--57.

\bibitem{CNSW2000prl}
D.S. Callaway, M.E.J. Newman, S.H. Strogatz, D.J. Watts, 
Network robustness and fragility: Percolation on random graphs, 
Phys. Rev. Lett. 85 (25) (2000) 5468--5471.

\bibitem{HZ2012g}
W.G. Hill, X.-S. Zhang, 
On the pleiotropic structure of the genotype--phenotype map and the evolvability of complex organisms, 
Genetics 190 (3) (2012) 1131--1137.

\end{thebibliography}

\end{document}